\documentclass[traditabstract]{aa} 
\usepackage{graphicx}
\usepackage{natbib}
\usepackage{txfonts}
\begin{document}
\title{The global dust SED: \\Tracing the nature and evolution of dust with {\tt DustEM} }

   \subtitle{}

   \author{M. Compi\`egne\inst{1}
\and
L. Verstraete\inst{2}
\and 
A. Jones\inst{2}
\and
J.-P. Bernard\inst{3}
\and
F. Boulanger\inst{2}
\and
N. Flagey\inst{4}
\and
J. Le Bourlot\inst{5}
\and
D. Paradis\inst{4}
\and
N. Ysard\inst{6}
}

\institute{Canadian Institute for Theoretical Astrophysics, University of Toronto, 
60 St. George Street, Toronto, ON M5S 3H8, Canada\\
\email{compiegne@cita.utoronto.ca}
\and
Institut d'Astrophysique Spatiale, UMR8617, CNRS, Universit\'e Paris-sud XI,
b\^atiment 121, F-91405 Orsay Cedex, France
\and
Centre d'Etude Spatiale des Rayonnements, CNRS et Universit\'e Paul Sabatier-Toulouse 3, Observatoire Midi-Pyr\'en\'ees,
 9 Av. du Colonel Roche, 31028 Toulouse Cedex 04, France
\and
Spitzer Science Center, California Institute of Technology, 1200 East California Boulevard, MC 220-6, Pasadena, CA 91125, USA
\and
Observatoire de Paris, LUTH and Universit\'e Denis Diderot, Place J. Janssen, 92190 Meudon, France
\and
Helsinki University Observatory, 00014, University of Helsinki, Finland
}

   \date{Received ; accepted }

 
   \abstract 
{The Planck and Herschel missions are currently measuring the far-infrared to millimeter emission of dust, which combined with existing IR data, will for the first time provide the full spectral energy distribution (SED) of the galactic interstellar medium dust emission, from the mid-IR to the mm range, with an unprecedented sensitivity and down to spatial scales $\sim30\arcsec$.  Such a global SED will allow a systematic study of the dust evolution processes (e.g. grain growth or fragmentation) that directly affect the SED because they redistribute the dust mass among the observed grain sizes. The dust SED is also affected by variations of the radiation field intensity.  Here we present a versatile numerical tool, {\tt DustEM} , that predicts the emission and extinction of dust grains given their size distribution and their optical and thermal properties.  In order to model dust evolution, {\tt DustEM}  has been designed to deal with a variety of grain types, structures and size distributions and to be able to easily include new dust physics. We use {\tt DustEM}  to model the dust SED and extinction in the diffuse interstellar medium at high-galactic latitude (DHGL), a natural reference SED that will allow us to study dust evolution. 
We present a coherent set of observations for the DHGL SED, which has been obtained by correlating the IR and HI-21 cm data. The dust components in our DHGL model are (i)\,polycyclic aromatic hydrocarbons, (ii)\,amorphous carbon and (iii)\,amorphous silicates. We use amorphous carbon dust, rather than graphite, because it better explains the observed high abundances of gas-phase carbon in shocked regions of the interstellar medium. Using the {\tt DustEM}  model, we illustrate how, in the optically thin limit, the IRAS/Planck\,HFI (and likewise Spitzer/Herschel for smaller spatial scales) photometric band ratios of the dust SED can disentangle the influence of the exciting radiation field intensity and constrain the abundance of small grains ($a\la10$ nm) relative to the larger grains. We also discuss the contributions of the different grain populations to the IRAS, Planck (and similarly to Herschel) channels. Such information is required to enable a study of the evolution of dust as well as to systematically extract the dust thermal emission from CMB data and to analyze the emission in the Planck polarized channels. The {\tt DustEM}  code described in this paper is publically available.}


\maketitle

\section{Introduction}
\label{sect:intro}

Dust plays a key role in the physics (e.g. heating of the gas, coupling to the magnetic field) and chemistry (formation of H$_2$, shielding of molecules from dissociative radiation) of the interstellar medium (ISM). Heated by stellar photons, dust grains radiate away the absorbed energy by emission in the near-IR to mm range. Dust emission can thus be used as a tracer of the radiation field intensity and, hence, of star formation activity. Assuming a constant dust abundance, the far-IR to mm dust emission is also used to derive the total column density along a line of sight and to provide mass estimates.  The impact of dust on the ISM and the use of its emission as a tracer of the local conditions depends on the dust properties and abundances.  It is therefore of major importance to understand dust properties and their evolution throughout the ISM. 

The instruments onboard the Herschel and Planck satellites are currently observing the dust spectral energy distribution (SED) from the FIR to the millimetre with unprecedented sensitivity and angular resolution. Complementing these observations with the existing IR data from ISO, Spitzer and IRAS, we will soon have at our disposal the global dust SEDs, where the emission from grains of all sizes is observed. Such data will allow us to undertake systematic studies of the key dust evolution processes in the ISM (e.g. grain growth and fragmentation), which are primarily reflected in the variation of the dust size distribution and, in particular, in the ratio of the abundance small grains ($a\la 10$ nm) to large grains \citep[e.g.][]{stepnik2003, paradis2009b, rapacioli2005, berne2007}.  
Dust evolution encompasses many complex processes and its impact may be studied using the dust SED for the diffuse interstellar medium at high galactic latitude (DHGL) as a reference. We present here a coherent data set for the DHGL and the corresponding dust model used to match these observations.

\begin{table*}[t]
\caption{Dust SED, $\nu\,I_{\nu}$, and its uncertainty, $\sigma(\nu\,I_{\nu})$, in unit of $10^{-30}\,W\,sr^{-1}\,H^{-1}$ measured for the Diffuse High Galactic Latitude (DHGL) medium.
  The DIRBE and WMAP values were obtained by correlation of the measured intensities with H~I emission data and corrected to account for the contributions of the diffuse ionized medium and the diffuse molecular gas (see \S\,\ref{sect:reference_sed}).
  The Herschel PACS, SPIRE and Planck/HFI values were derived by integrating the FIRAS spectrum  (see \S\,\ref{sect:reference_sed}) considering the instrument transmission and color correction.}
  \centering
 \begin{tabular}{ c c c c c c c c c c c }
   \hline           
   \hline   
    Instrument    &DIRBE & DIRBE & DIRBE & DIRBE & DIRBE& DIRBE & DIRBE & DIRBE & WMAP & WMAP\\
       $\lambda\,(\mu m)$  &  3.5 & 4.9 & 12  & 25 &  60  & 100 & 140 & 240 & 3200 & 4900 \\
   \hline           
$\nu\,I_{\nu}$              & 8.4 & 9.5 & 61.6 & 30.7 & 47.6 & 160.6 & 216.6 & 99.2 & 4.0\,$10^{-3}$ & 8.2\,$10^{-4}$\\
$\sigma(\nu\,I_{\nu})$ & 1.0 & 1.1 & 7.3 & 3.6 & 1.1 & 19.0 & 18.8 & 8.9 & 1.5\,$10^{-4}$ & 4.9\,$10^{-5}$\\
\hline
 & & & & & & & & & & \\
  \cline{2-9}   
     & Instrument &    PACS & SPIRE & SPIRE  & SPIRE & HFI & HFI & HFI &   & \\ 
    & $\lambda\,(\mu m)$    & 160 & 250 & 350 & 500 & 350 & 550 & 850  & & \\
   \cline{2-9}   
& $\nu\,I_{\nu}$  &   217.0 & 100.0 & 37.3 & 11.6 & 36.0 & 8.3 & 1.4 & &  \\
& $\sigma(\nu\,I_{\nu})$   & 34.7 & 9.3 & 3.3 & 1.5 & 3.2 & 1.4 & 0.3 & &   \\
    \cline{2-9}        
\end{tabular}
 \label{tab:obs_DHGL} 
\end{table*}

A suitable dust model has to, ideally, explain all of the available observational constraints, these are primarily: (i) the extinction curve, (ii) the emission brightness spectrum $\nu I_{\nu}$ or SED (Spectral Energy Distribution), (iii) the elemental depletions, (iv) the spectral dependence and efficiency of polarization in emission and extinction and finally, (v) the plausibility of a given dust component in view of its likely formation and destruction in the ISM. 
Early models focused on explaining the extinction (some of them also describing polarization by preferential extinction) and used silicates and carbonaceous material \citep[e.g.][]{MRN1977, draine84, kim94a, kim95, li97}. 
In the 1980's, the spectroscopy of dust emission in the IR lead to the introduction of a population of aromatic hydrocarbons called polycyclic aromatic hydrocarbons (PAHs).  However, current models aim to consistently reproduce the observed extinction and emission \citep[][hereafter DBP90]{desert90} \citep[][hereafter DL07]{siebenmorgen92, dwek97,li2001,zubko2004,draine2007} as well as the associated polarization \citep[][]{draine2009}.
Further details of the dust modelling process can be found in recent reviews on interstellar dust, such as those by  \citet{Draine2003b, li2003, draine2004}.  
We emphasize here that all of the different dust models, developed to date, have been designed using essentially the same observational constraints \citep[e.g.][]{li97,zubko2004,draine2007}.

Carbonaceous dust is often considered to be in the form of crystalline graphite because it can explain the 217\,nm bump in the observed extinction curve.  The question of the origin of this feature is however still open because graphite cannot explain the band profile variability observed in the ISM \citep{fitzpatrick2007,Draine2003b}. In fact, throughout the dust lifecycle, from dense molecular cores to the diffuse ISM, the grain populations are the result of complex, non-equilibrium evolutionary processes and it appears natural to consider that carbon dust is amorphous, as observations have clearly demonstrated for silicate grains \citep[e.g.][]{li2001a, kemper2005, li2007}.  We follow this train of thought in this work and examine the possibility that all carbonaceous grains (except PAH) are amorphous and partially hydrogenated in the interstellar medium.

The paper is organized as follows: we first present our model {\tt DustEM}  and the developed dust SED-fitting tool  (Sect\,\ref{sect:skills}). We then discuss why amorphous carbon is the most likely carbonaceous dust grain component in the ISM (Sect\,\ref{sect:graph_free}).  Section\,\ref{sect:DHGL} presents a coherent data set for the DHGL SED and the associated dust model that serves as a reference SED for the evaluation of the nature and efficiency of dust evolution processes. In Sect.\,\ref{sect:evolution}, we show how changes the dust abundances and sizes can be traced from their global SED, thus allowing a study of dust evolution.  The conclusions and perspectives of this work are drawn together in Sect.\ref{sect:summary}.

\section{The {\tt DustEM}  model capabilities}
\label{sect:skills}
Accounting for the emission from dust, for which the properties evolve in the ISM, requires a model that easily allows the dust properties to be changed and mixed, and where new dust physics can easily be incorporated and tested. In this section we present  a numerical tool, {\tt DustEM} , which is designed to this end and that provides the extinction and emission for interstellar dust populations.  Note that {\tt DustEM}  computes the local dust emissivity, and as such, does not include any radiative transfer calculation, so that it can be included as a source and extinction term in more complex applications explicitly dealing with energy transport through optically thick media.

\subsection{The {\tt DustEM}  model}
\label{sect:dustem}
The {\tt DustEM}  model stems from that of DBP90 but the code has been completely re-written in fortran 95 in order to achieve the goals described above. We use the formalism of \citet{desert86} to derive the grain temperature distribution $dP/dT$ and {\tt DustEM}  then computes the dust SED and associated extinction for given dust types and size distributions. To correctly describe the dust emission at long wavelengths the original algorithm has been adapted to better cover the low temperature $dP/dT$ region. Using an adaptative temperature grid, we iteratively solve Eq. 25 from \citet{desert86} in the approximation where the grain cooling is fully continuous\footnote{The grain cooling is taken to be as in Eq. 8 of \citet{desert86} where the heating term is neglected and the cut-off energy is that of the hardest photons in the radiation field.}. This topic will be discussed in detail in a forthcoming paper, along with its consequences for the emission of small grains and gas-grain interactions. 
Note that unless explicitly specified in the input file, the dP/dT calculation is performed for 
all grain populations and sizes including those for which the thermal equilibrium approximation would apply.
The principal input parameters for the dust model (e.g. grain types, dust-to-gas mass ratios) used to compute the required quantities are read from a control input file along with keywords (e.g. that specify the size distributions or the outputs).  {\tt DustEM}  can handle an arbitrary number of grain types. The built-in size distributions $dn/da$ (the number of grains of radius between $a$ and $a+da$) are either power law or log-normal but any size distributions can be given in input data files. For each grain of specific mass density $\rho$, the radius $a$ is defined as that of an equivalent sphere of mass $m=\rho\,4\pi a^3/3$.  For a given dust type, the optical  properties are provided as a function of the grain radius $a$ in the data files, namely: $Q$-files for the absorption and scattering efficiencies, $Q(a,\lambda)$, and $C$-files for the heat capacity per unit volume, $C(a,T)$\footnote{Size-dependent heat capacities are required to describe PAH (C-to-H ratio) and composite grains.}. The $Q$ and $C$ values are provided over the broadest possible size range (usually $\sim$0.3 to $10^4$ nm). Any  grain type can be considered once these $Q$ and $C$-files have been generated and are in the {\tt DustEM}  data file set. The requested dust sizes have to fit within those used in the above files (i.e. extrapolation is not allowed).

A size-dependent weight factor (or mixing) read from a data file can be applied to any grain type while computing its extinction and emission. This allows us to take into account a number of effects, e.g., the ionization of PAHs or more generally the evolution of one grain type to another as a function of size.  The format of the {\tt DustEM}  tabulated size distributions (and their elemental composition as a function of size, in case of composite grains) data files is shared with {\tt DustEV}, the dust evolution model described in Guillet et al. (2010, in preparation) and which follows the evolution in composition and size of charged dust grains as a result of coagulation and fragmentation processes.  The default outputs from {\tt DustEM}  are the emissivity per hydrogen atom ($N_{\rm H}\,=\,N_{\rm HI}\,+\,2\,N_{\rm H_2}$) for each grain type $4\pi\,\nu I_{\nu}/N_{\rm H}$ in erg\,s$^{-1}$\,H$^{-1}$ and the extinction cross-section per gram for each grain type (in cm$^2$\,g$^{-1}$). When required (and via keywords in the input control file) the grain temperature distribution and the dust emissivity can be provided for each grain size. When required {\tt DustEM}  also allows the use of temperature-dependent dust emissivity (see Appendix\,\ref{sect:qabstemp}).

The current development plan for {\tt DustEM}  consists of the inclusion of (i) the spinning dust emission, (ii) the polarized extinction/emission, and (iii) the physics of amorphous solids at low temperatures \citep{meny2007}. As for the mixing or the $\beta$-correction, each process is activated with a keyword given in the input control file and the associated parameters or tabulated values are then given in a specific file. {\tt DustEM}  can also be used as a subroutine in radiative transfer models, such as the Meudon PDR code\footnote{{\tiny Available at {\tt http://pdr.obspm.fr/PDRcode.html}}}, which solves for the gas state and dust emissivity. {\tt DustEM}  can also be used as a subroutine of the 3D-continuum transfer model (CRT)\footnote{{\tiny Available at \\
{\tt http://wiki.helsinki.fi/display/$\sim$mjuvela@helsinki.fi/CRT}}} of \citet{juvela2003}.

\begin{table}
\caption{DHGL dust model abundances and size distribution parameters  (see\,\S\ref{sect:reference_model}).
               Y is the mass abundance per hydrogen for each dust component. $f_{M_{tot}}$ is the dust component mass as a fraction of the total dust mass.
         }
  \centering
  \begin{tabular}{ c c c c c c c }
   \hline           
   \hline           
                                & $\sigma$ & $a_0$ &   & &   Y  & $f_{M_{tot}}$ \\
                               &                  &   $(nm)$     &     &   & $(M/M_H)$   & \\
  \hline
  PAH                        &  0.1      &   0.64 &    & & $7.8\,10^{-4}$    &  7.7\%  \\
  SamC                     &  0.35     &    2.0   &  &     &   $1.65\,10^{-4}$  &  1.6\%\\
                               &        &      &  &     &     &  \\
\hline
                                & $\alpha $ & $a\rm{_{min}}$ & $a_c, a_t$  & $\gamma$ &  \\
                               &         &     $(nm)$     &    $(nm)$   &   &    & \\
\hline
  LamC                    &   -2.8    & 4.0   &  150   & 2.0  &  $1.45\,10^{-3}$   & 14.2\% \\
   aSil                       &  -3.4    &  4.0   &  200  &  2.0 &   $7.8\,10^{-3}$  &  76.5\% \\ 
 \hline
                               &        &      &  &   TOTAL  &  $10.2\,10^{-3}$   &  \\
  \hline
 \end{tabular}
  \label{tab:model_param} 
\end{table}
\begin{figure}
    \centering
      \includegraphics[width=0.49\textwidth]{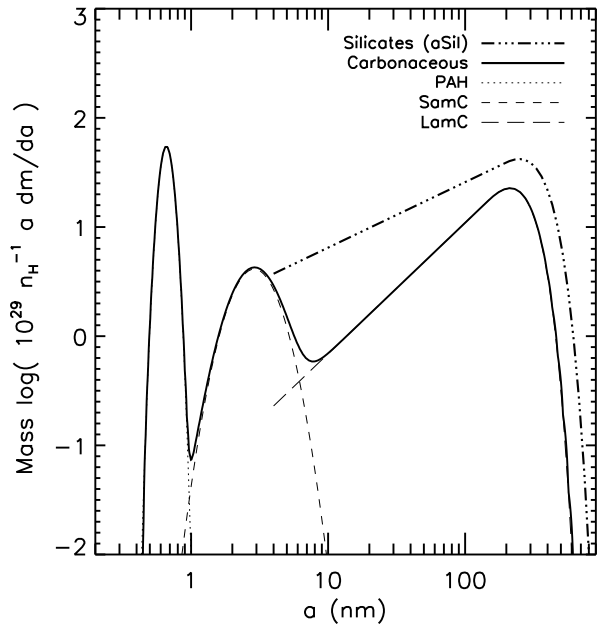}
     \caption{Mass distribution for the four dust types used in the DHGL model.}
          \label{fig:sdist}
 \end{figure}

\begin{figure*}[t]
    \centering
     \includegraphics[width=0.9\textwidth]{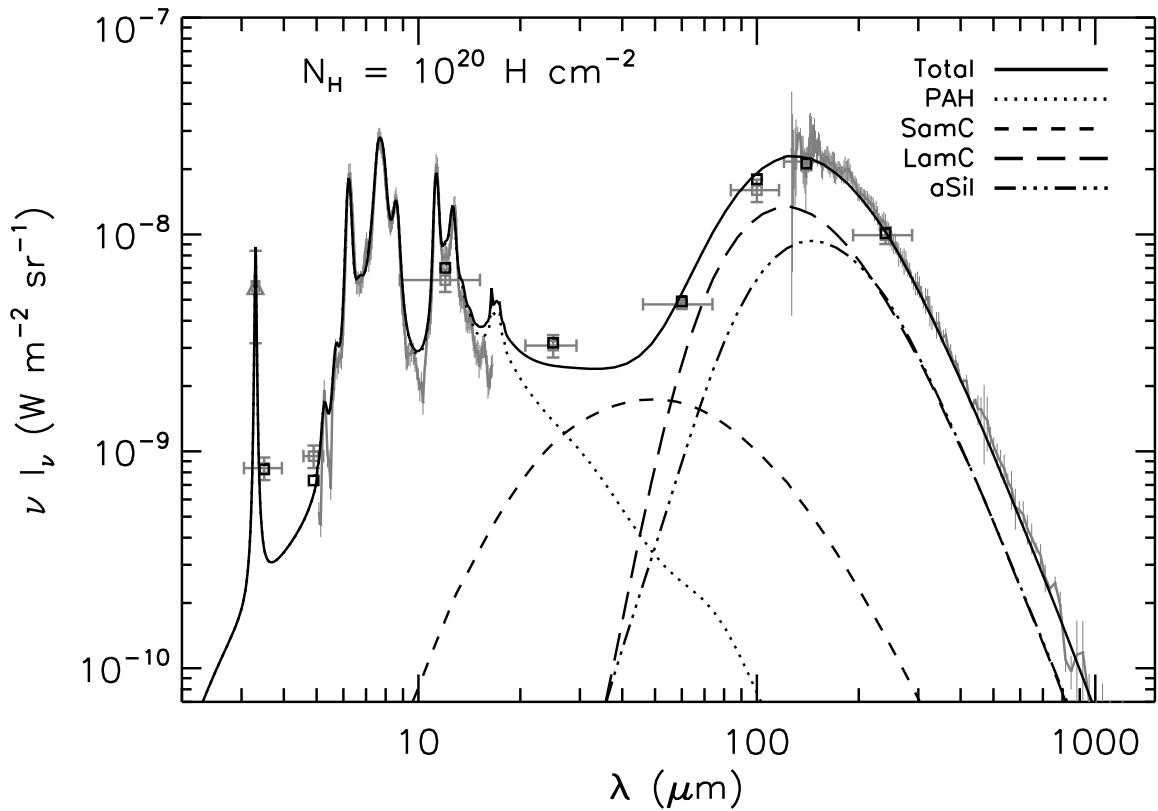}
      \caption{
       Dust emission for the DHGL medium.  
        Grey symbols and curves indicate the observed emission spectrum (see
        \S\,\ref{sect:reference_sed}) for $N_H\,=\,10^{20}$ H cm$^{-2}$.  
        The mid-IR ($\rm{\sim5-15\,\mu m}$) and far-IR ($\rm{\sim100-1000\,\mu m}$)
        spectra are from ISOCAM/CVF (ISO satellite) and FIRAS (COBE satellite), respectively.
       The triangle at 3.3 $\mu$m is a narrow band measurment from AROME balloon experiment.
      Squares are the photometric measurments from DIRBE (COBE).
Black lines are the model output
         and black squares the modeled DIRBE points taking into account instrumental transmission and color corrections.
}
          \label{fig:modelemiss}
\end{figure*}

\subsection{The IDL wrapper}
The {\tt DustEM}  code is coupled to a wrapper written in IDL (Interactive Data Language). The wrapper allows the user to run {\tt DustEM}  iteratively from a code development and debugging environment.  In particular, it is possible through this interface, to provide a given SED, and the associated error bars, to be fitted by {\tt DustEM} . The SED fitting uses the {\sc mpfit} \citep{markwardt2009} IDL minimization routine\footnote{{\tiny To be found at {\tt http://purl.com/net/mpfit}}}, which is based on the Levenberg-Marquardt minimization method. This allows the user to iteratively modify any of the model's free parameters, in order for the model prediction to converge on the provided SED. In this process the wrapper uses the {\tt DustEM}  computed spectrum to predict the SED values that would be observed by astronomical instruments, iteratively taking into account the appropriate color correction for wide filters and the flux conventions used by the instruments. The currently recognized settings include any filter used by the IRAS, COBE (DIRBE, DMR), Spitzer (IRAC, MIPS), Planck (HFI, LFI) and Herschel (PACS, SPIRE) satellite instruments, as well as that of the Archeops and Pronaos balloon-borne experiments.  The target SED can also include arbitrary spectroscopic data points, (such as for the COBE-FIRAS or Spitzer-IRS observations) for which no color correction is applied.  The fitting of an SED to a reasonable accuracy through this interface generally requires a few minutes of a single CPU.  The resulting best fit parameters are provided with their 1-sigma error bars, as derived from the {\sc mpfit} routine.  The wrapper also uses the capability of {\sc mpfit} to tie various free parameters through a user specified function.  It also includes the concept of user-written subroutines (e.g. a subroutine to manage the PAH ionization fraction).  The wrapper can also be used to generate pre-calculated SED tables for the instruments mentioned above, which can then be used to perform regression for a limited set of free parameters. This method is well adapted to problems where the direct fitting of individual SEDs over many image pixels would require too much CPU time.

Past applications of the {\tt DustEM}  wrapper to fit SEDs have included deriving the InterStellar Radiation Field (ISRF) scaling factor and the dust mass abundances (Y) for the various dust grain types, as well as constraining the size distribution of very small grains (VSGs), based on SEDs including a combination of IRAS and Spitzer data \citep[e.g.][]{bernard2008, paradis2009a} and Herschel data (Bernard et al. 2010).  Future developments of the wrapper will include the possibility to simultaneously fit extinction and emission and to follow the planned evolution of the fortran code.

{\tt DustEM}  is developed within the framework of a collaboration between IAS and CESR. The fortran code, the IDL wrapper and the associated data files can be downloaded from {\tt http://www.ias.u-psud.fr/DUSTEM}.\\

\section{The nature of interstellar carbon grains}
\label{sect:graph_free}

\begin{figure}
    \centering
      \includegraphics[width=0.49\textwidth]{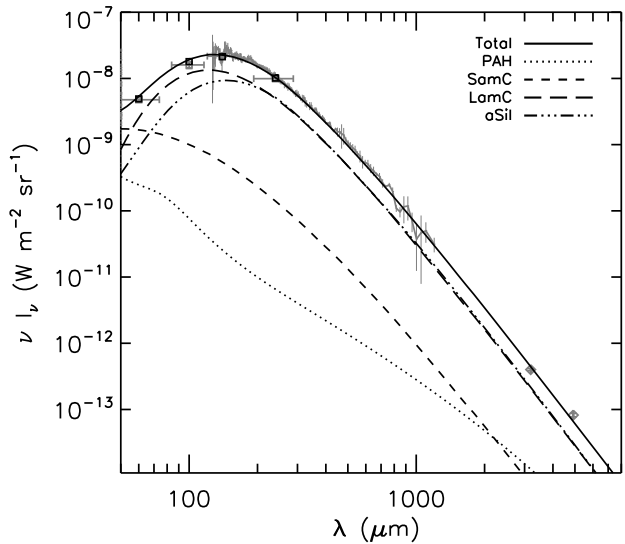}
     \caption{
Long wavelength emission of dust in the DHGL medium.
       Grey symbols and curves indicate the observed emission spectrum (see
        \S\,\ref{sect:reference_sed}) for $N_H\,=\,10^{20}$ H cm$^{-2}$.  
      The spectrum is a FIRAS (COBE satellite) measurment.
     Squares and diamonds are the photometric measurments from DIRBE (COBE) and WMAP, respectively.
Black lines are the model output  and black squares the modeled DIRBE points taking into account instrumental transmission and color corrections.
         }
          \label{fig:modelemiss_lw}
\end{figure}

 \begin{figure}
    \centering
      \includegraphics[width=0.49\textwidth]{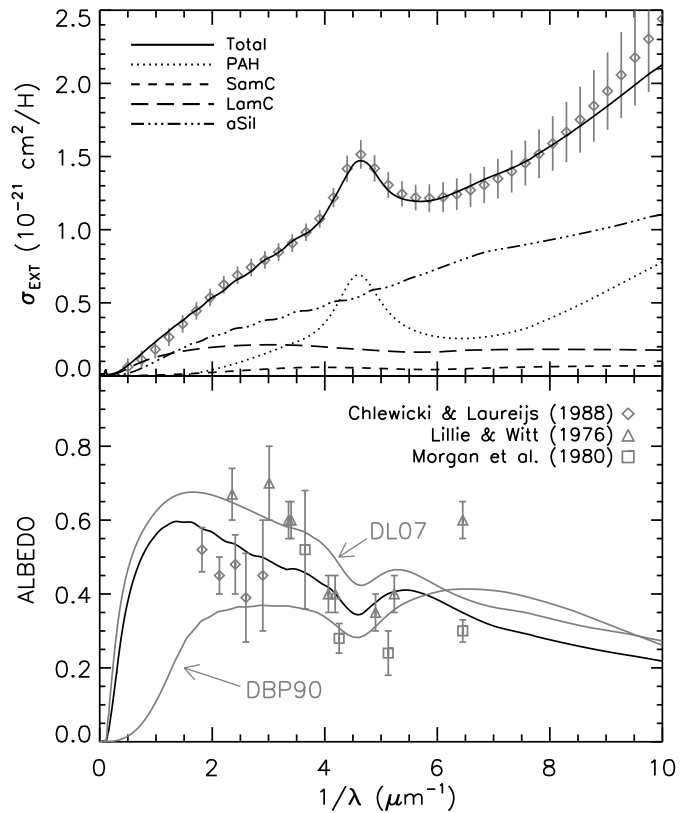}
     \caption{The extinction curve (top) and albedo (bottom) for dust in the
        DHGL medium.  In the top panel, the grey diamonds
        show the \citet{fitzpatrick99} extinction law for
        $R_V\,=\,3.1$. Error bars come from the dispersion of the observed extinction curve.
        The bottom panel overlays the model albedos from DBP90 and DL07 (grey lines) 
        and from the {\tt DustEM}  model (black line), compared to the observational data points.}
          \label{fig:modelextUV}
 \end{figure}
\begin{figure}
    \centering
      \includegraphics[width=0.49\textwidth]{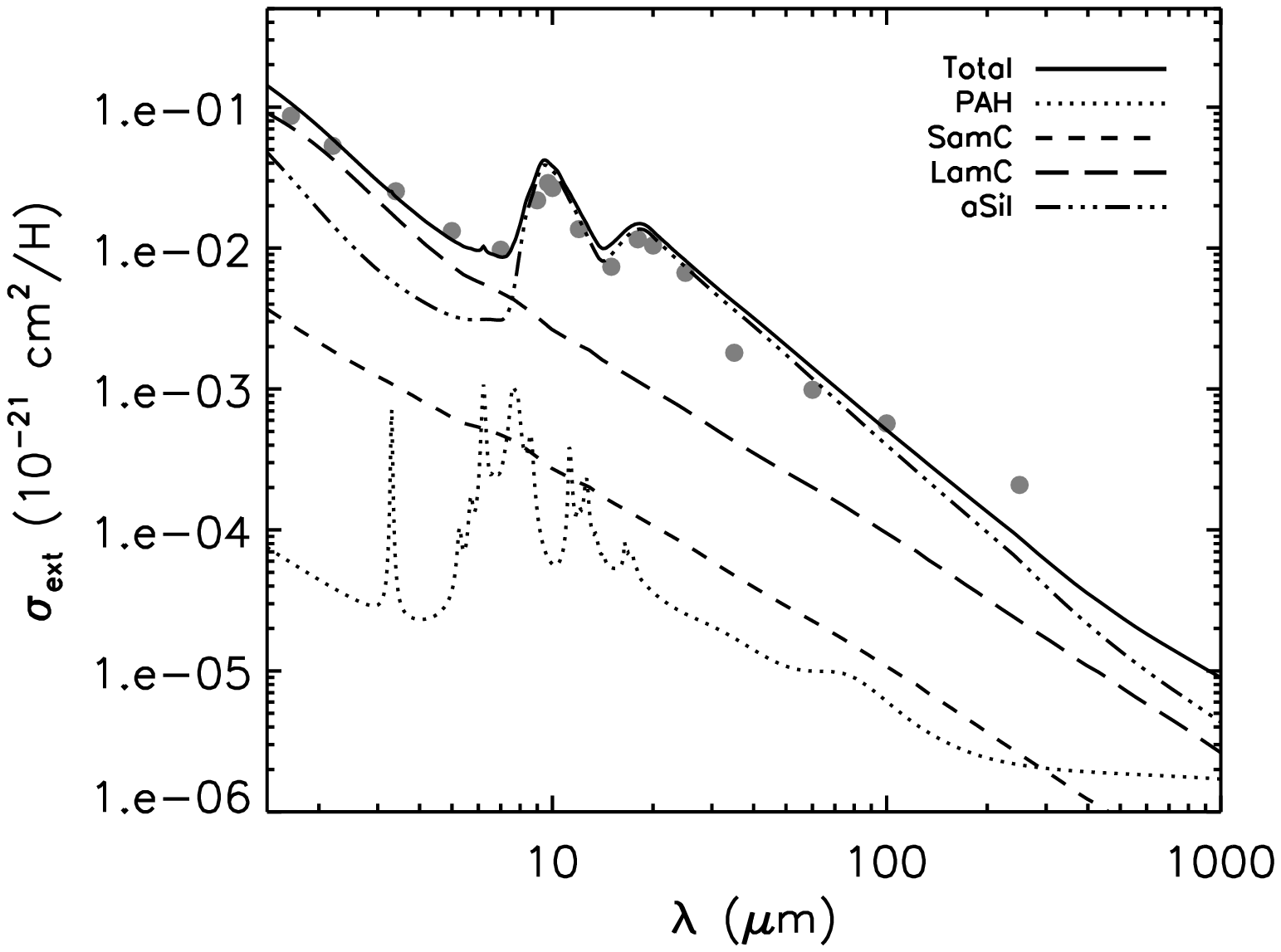}
     \caption{Infrared extinction of dust in the
        DHGL medium. The grey
        circles show the observed extinction \citep{mathis90}. Black lines are the model.}
          \label{fig:modelext}
\end{figure}

We discuss here the plausible nature of carbon dust grains in the ISM.  Carbonaceous dust is formed in the ejecta of evolved stars but, seemingly, must also form directly in the ISM \citep{zhukovska2008, draine2009a, jones2009b}.  The smallest carbonaceous species contain a few tens of C-atoms and are assigned to PAH-like species due to their spectral similarity to the observed mid-IR emission bands. The nature of the larger carbonaceous grains is less clear, in particular, because of the lack of specific spectroscopic signatures.  Graphite was long-ago
proposed as an interstellar dust component because its optical anisotropy could naturally explain the polarization of starlight dimmed by dust \citep[e.g.][]{cayrel54} and especially since it features a strong absorption band (due to $\pi\rightarrow \pi^*$ transitions in sp$^2$ carbon) similar to the 217\,nm bump observed in interstellar extinction \citep{stecher65}.
However, such a feature exists in many carbonaceous materials containing aromatic rings and is not specific to graphite. In addition, detailed modelling of the band profile by \citet{draine93b} has shown that graphite grains cannot explain the observed bandwidth variability \citep{fitzpatrick2007}. \citet{draine93b} suggest that this variability may be due to the presence of hydrogen or defects within graphite.

In fact, observational evidence strongly suggests that carbon dust consists of hydrogenated amorphous carbon (HAC or a-C:H) materials that form around evolved stars, metamorphose there \citep[e.g.,][]{goto2003} and further evolve in the ISM \citep[e.g.,][]{jones90,pino2008,jones2009b}.  
HAC material could also form directly into the ISM from the UV processing of ice mantles after leaving
the dense medium as suggested by \citet{greenberg95} or via direct accretion of carbon and hydrogen \citep[e.g.][]{jones90}.
In quiescent regions of the diffuse ISM $\sim 40$\% of the available carbon is observed to be in the gas \citep[e.g.,][]{cardelli96}. In contrast, in regions of the ISM shocked to velocities of the order of $100-150$\,km\,s$^{-1}$ as much as $80-100$\% of the carbon atoms are observed to be in the gas phase \citep{frisch99,welty2002, podio2006,slavin2008}, indicating that a large fraction of the carbon dust has been destroyed.  Graphite/amorphous carbon dust processing models \citep{tielens94,jones94,jones96} predict that, at most, only $15$\% of graphitic carbon dust should be eroded in such shocks (for silicate dust the observations and model results agree rather well). This discrepancy for carbon dust implies that we need to re-consider the nature of carbonaceous dust in the ISM. \citet{serra2008} re-evaluated the processing of carbon grains in shocks, using the physical parameters typical of a-C:H/HAC, rather than of graphite. Their newly-calculated sputtering yields for a-C:H/HAC indicate that it is significantly more susceptible to sputtering than graphite and they showed that $30-100$\% of these type of carbonaceous grains are destroyed ($\equiv 60-100$\% carbon in the gas) in $100-200$\,km\,s$^{-1}$ supernova-generated shock waves in the warm inter-cloud medium. These results, along with similar results obtained for PAH evolution in shocks and in a hot gas \citep{micelotta2010a,micelotta2010b}, are in good agreement with the observations of shocked regions of the ISM \citep[e.g., $80-100$\% of the carbon in the gas;][]{frisch99,welty2002,podio2006,slavin2008}.  

Thus, even if all the carbonaceous dust formed around evolved stars were to be in graphitic form, an unlikely scenario, it would subsequently be modified, by ion irradiation \citep[e.g.,][]{banhart99,mennella2003,mennella2006} in shocks and by cosmic rays in the ISM, to an amorphous form implanted with heteroatoms, principally H atoms.  However, given the results for the processing of a-C:H/HAC dust and PAHs \citep{serra2008,micelotta2010a,micelotta2010b}, the ion and electron irradiation of carbon grains in shocks appears to be very destructive. Therefore, the carbon dust in the ISM is probably completely re-processed on short time-scales, rather than simply being modified by ion irradiation effects in shocks.

The physical and chemical properties of a-C:H/HAC solids are sensitive to the external conditions (e.g., photon, ion or electron irradiation) and can undergo a process of `aromatization' when heated or exposed to UV or ion irradiation \citep[e.g.,][]{duley89,jones90b,jones90,jones2009b}.  Laboratory experiments on interstellar carbon dust analogues \citep[e.g.,][]{dartois2004} show that hydrogen-rich ($>$ 50 atomic \% H) HAC solids can explain the interstellar absorption bands at 3.4, 6.85 and 7.25~$\mu$m and that the thermal annealing of the material is accompanied by an increase in the aromatic carbon content  (i.e., `aromatization').  The HAC model for hydrocarbon grains in the ISM \citep[e.g.,][and references therein]{jones2009b} would therefore predict that the larger grains, with temperatures in equilibrium with the local radiation field, should be rather hydrogen-rich because they do not undergo significant heating. In contrast, the smaller, stochastically-heated grains, will be converted to hydrogen-poorer, lower density, smaller bandgap, aromatic-rich materials. The photo-fragmentation of the smaller, aromatic-rich grains could then be an important source of molecular aromatic species such as PAH s and molecular fragments \citep[e.g.,][]{jones90b,duley2000,petrie2003,pety2005,jones2005, jones2009b}.

 In addition to the PAHs, we assume that the interstellar carbonaceous grains are in the form of hydrogenated amorphous carbons, collectively known as a-C:H or HAC, rather than graphite, and adopt these materials as an analogue for interstellar carbon dust \citep[e.g.,][]{jones90b,dartois2004,dartois2007a,dartois2007,pino2008,jones2009b}. For the optical properties of these  amorphous carbons, we here use the BE sample complex refractive index data, derived from laboratory measurement by \citet{zubko96a}, to be representative of sp$^2$ and sp$^3$ containing (hydrogenated) amorphous carbons (see Appendix\,\ref{sect:amC}). However, we point out that these materials can have a wide range of properties and these can evolve as a function of the local conditions.  

\begin{figure}[t]
   \centering
     \includegraphics[width=0.49\textwidth]{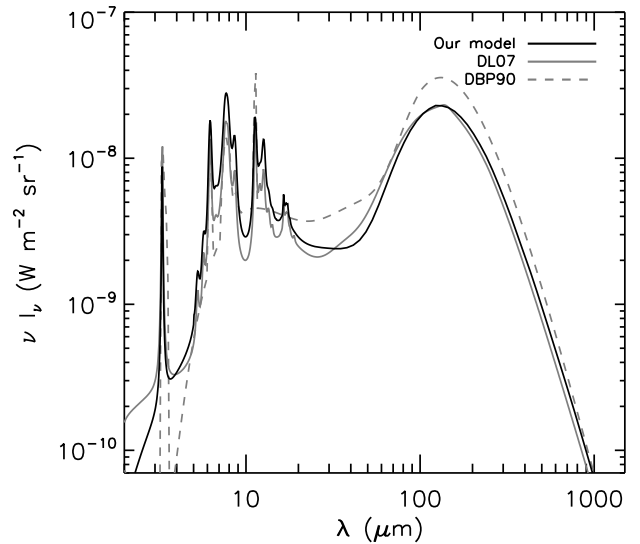}
   \caption{The dust emission for our DHGL model compared to those of DL07 (U=1, $q_{PAH}\,=\,4.6\,\%$) and DBP90 for $N_H\,=\,10^{20}$ H cm$^{-2}$. 
                 For DBP90, the measured SED used as a reference was higher explaining the deviation at wavelengths larger than $\sim 20\;\mu$m.}
          \label{fig:compare_emiss}
\end{figure}
 
\section{A dust model for the diffuse high galactic latitude medium}
\label{sect:DHGL}

We follow earlier studies (e.g. DPB90) in using observations of the diffuse interstellar medium in the Solar Neighborhood to constrain our dust model. The model fitted to these observations defines a reference SED with which to compare and characterize dust evolution as a function of the local interstellar environment in the Galaxy and in external galaxies. In this section, we describe the observations used and the model that reproduces these observations.

\begin{table*}[t]
 \begin{center}
    \caption{
        The elemental abundances (see \S\,\ref{sect:reference_sed} for details).
}
  \begin{tabular}{l c c c c c}   
 \hline                 
                              &   C   &   O     &  Mg   &    Si    &   Fe    \\
 \hline
    Measurements & &  &  & &  \\
  \hline
  \tablefootmark{a}$\rm{[X/10^6H]_{\sun}}$         & 269$\pm$33 & 490$\pm$60 & 40$\pm$4 & 32$\pm$2 & 32$\pm$3 \\
 \tablefootmark{b}$\rm{[X/10^6H]_{F,G\star}}$      & 358$\pm$82 &445$\pm$156  & 42.7$\pm$17.2 & 39.9$\pm$13.1 & 27.9$\pm$7.7\\
\hline
  \tablefootmark{c}$\rm{[X/10^6H]_{gas}}$            & 75$\pm$25 & 319$\pm$14 & $\sim$0 & $\sim$0  & $\sim$0 \\
\hline
  $\rm{[X/10^6H]_{dust/\sun}}$    & 194$\pm$41 & 171$\pm$62 & 40$\pm$4 & 32$\pm$2 & 32$\pm$3 \\
 $\rm{[X/10^6H]_{dust/F,G\star}}$ & 283$\pm$86 & 126$\pm$157  & 42.7$\pm$17.2 & 39.9$\pm$13.1 & 27.9$\pm$7.7\\
  \hline
    DHGL dust model & &  &  & &  \\
    \hline
   $\rm{[X/10^6H]_{PAH}}$        &  65   & - &  - & - & - \\
   $\rm{[X/10^6H]_{SamC}}$      &  14  & - &  - & - & - \\
   $\rm{[X/10^6H]_{LamC}}$      &  121  & - &  - & - & - \\
   $\rm{[X/10^6H]_{aSil}}$        &  -  & 180 & 45  & 45 & 45 \\
    \hline
    $\rm{[X/10^6H]_{TOTAL}}$   & 200 & 180 & 45  & 45 & 45 \\
  \hline                    
\end{tabular}
\tablefoot{
\tablefoottext{a}{from \citet{asplund2009}}\\
\tablefoottext{b}{from \citet{sofia2001}}\\
\tablefoottext{c}{C estimate is from \citet{dwek97} while the O estimate is from \citet{meyer98}}\\
}
    \label{tab:abundances}   
\end{center}
 \end{table*}

\subsection{The observables}
\label{sect:reference_sed}

A first observational constraint on the dust model comes from measurements of the dust extinction and scattering properties. For the extinction curve we use the data, for $R_V=3.1$, from \citet{fitzpatrick99} in the UV-visible and  \citet{mathis90} in the infrared. We normalize the observations using the empirical determination of $N_H/E(B\,-\,V)\,=\,5.8\times10^{21}$ H cm$^{-2}$, obtained by correlating the reddening of stars with the HI and H$_2$ column densities derived from Copernicus and FUSE far-UV spectroscopic observations \citep{bohlin78,rachford2002}.  The dust albedo is taken from the work of \citet{lillie76, morgan80, chlewicki88}.

The dust model is also constrained by observations of the dust emission from near-IR to mm wavelengths. We use data at high Galactic latitudes from the COBE and WMAP space missions to determine the SED of the dust emission.  Several authors \citep[e.g.][]{boulanger96,arendt98} have analyzed the COBE data, but, to our knowledge, no publication presents a coherent set of measurements obtained over the same sky area. To reduce the uncertainties associated with variations of the dust emission per Hydrogen atom across the high Galactic latitude sky, we have made our own determinations. We used the DIRBE and FIRAS data sets, corrected for zodiacal light, available at {\tt lambda.gsfc.nasa.gov}.  For WMAP, we used the seven-year temperature maps \citep{jarosik2010}, and subtracted the cosmic microwave background (CMB) using the map obtained with the internal linear combination method \citep{gold2010}.  The free-free contribution to the emission was subtracted using the model of \citet{dickinson2003}.

At wavelengths $\ge 60\,\mu$m up to the WMAP centimetric data we determined the interstellar emission per Hydrogen atom by correlating the dust emission with the HI line emission, $I_{HI}$, taken from the Leiden-Argentina-Bonn survey \citep{kalberla2005}.  We obtained a coherent set of measurements by using the same sky area, defined by the galactic latitude and HI emission constraints $|b|>15^{\degr}$ and $I_{HI} \le 300$ K km s$^{-1}$ , for all wavelengths.  We discarded bright point sources, nearby molecular clouds in the Gould Belt, and the Magellanic clouds. The error-bars on each measurement were obtained by dividing the sky area into four sub-areas, and comparing results between sub-areas. These error-bars are dominated by systematic variations of the SED from one sub-area to the others.

It is not possible measure the interstellar emission at near- and mid-IR wavelengths over the same sky area that we used for the longer wavelengths.  At mid-IR wavelengths, and outside clean areas near the ecliptic poles, uncertainties in the zodiacal light removal are larger than the high galactic latitude interstellar emission. At near-IR wavelengths the stellar emission is dominant and its subtraction is an additional difficulty in measuring the interstellar emission. At wavelengths $\le 25\,\mu$m, we have thus used the colors, $I_\nu(\lambda)/I_\nu(100\, \mu$m), measured over distinct sky areas by \citet{arendt98} (see their Table 4).  We point out that their 140/$100\, \mu m$ and 240/$100\, \mu m$ colors, measured over a smaller sky area near the Galactic poles $\rm |b| >45^\circ$, are $10-15\%$ lower than our values. This difference reflects small variations of the far-IR colors across the high Galactic latitude sky. These spatial variations make the near and mid-IR parts of the SED more uncertain than the far-IR part, which we determined consistently over a single sky area.  We have complemented the DIRBE data points with the narrow band ($\Delta\lambda\,=\,0.17\,\mu$m) photometric point at 3.3\,$\mu$m from the AROME balloon observatory measured towards the molecular ring \citep{giard94}. The mid-IR spectrum is the ISOCAM/CVF spectrum for a diffuse region at galactic coordinates (28.6, +0.8) presented in \citet{flagey2006} and scaled to match the $12\,\mu$m emission per hydrogen.

For optically thin emission the HI cut-off used in the correlation analysis corresponds to a column density of neutral atomic hydrogen of $5.5 \times 10^{20}$ cm$^{-2}$.  Over this sky area, the interstellar emission arises primarily, but not exclusively, from dust within neutral atomic interstellar gas. Emission from dust associated with diffuse H$_2$ and HII gas also needs to be taken into account. Using the FUSE data reported by \citet{gillmon2006}, we estimated the fraction of H$_2$ gas over the selected sky area to be $\sim 3\%$.  Ionized gas accounts for a larger fraction \citep[$\sim 20\%$][]{reynolds89} of the hydrogen in the diffuse ISM. The H$_\alpha $ emission from the diffuse HII gas has been shown to be spatially correlated with HI gas emission \citep{reynolds95}. We assume that the dust emission per hydrogen atom is the same in the neutral and ionized components of the diffuse interstellar medium, and thus we scale the SED derived from the dust-HI correlation by 0.77, to obtain a dust emission per hydrogen atom that takes into account the contributions of the diffuse ionized medium and the diffuse molecular gas.  The corrected SED values and error-bars are listed in Table\,\ref{tab:obs_DHGL}, where the FIRAS spectrum has been convolved with the Herschel and Planck transmission for the relevant spectral bands. The full SED is plotted in Fig.\,\ref{fig:modelemiss} and Fig.\,\ref{fig:modelemiss_lw}.  From $100\,\mu m$ to 3\,mm, the SED is well fit by a black body spectrum with a temperature of 17.5~K, and an emissivity $\propto\nu^2$ for $\lambda \le 450\,\mu$m and $\propto\nu^{1.65}$ at longer wavelengths.  The diffuse HII must also contribute to the extinction along diffuse lines of sights as suggested by the difference between intercloud and cloud for the value of $(N(HI)+N(H_2))/E(B-V)$ reported by \citet{bohlin78}. Here we use the standard value of $5.8 \times 10^{21}$ cm$^{-2}$ mag$^{-1}$ also found by \citet{rachford2002} towards ``translucent lines of sight'' ($f_{H_2}\sim 0.5$, $1 \la A_V \la 3$).

Table\,\ref{tab:abundances} lists the measured elemental abundances and inferred dust elemental abundance relevant for our model.
The solar abundances are those of \citet{asplund2009} while the F, G stars (young $\le2$\,Gyr F, G disk stars) abundances are those of \citet{sofia2001}.
We do not consider B stars that have lower metallicity, because of the possible segregation of refractory elements 
during the star formation due to sedimentation and/or ambipolar diffusion processes \citep{snow2000}.
Note that this could also be the case for lower mass stars \citep[e.g. the sun, ][]{asplund2008}, raising the question of the existence of
a stellar standard that represents the overall ISM abundances.
The interstellar gas phase abundances used are from \citet{dwek97} for the carbon and from \citet{meyer98} for the oxygen.
The \citet{dwek97} value of 75$\pm$25\,ppm for the gas phase C abundance is in good agreement with
recent estimates by \citet{sofia2009}.
We emphasis that, as argued by \citet{draine2009b}, in view of the assumption that
are made for dust properties (especially density/porosity) and the uncertainties on the abundances, 
a model that departs from the measured elemental solid phase abundances by tens of percent should still be considered as viable.

\subsection{The model}
\label{sect:reference_model}

\begin{figure*}[t]
    \centering
      \includegraphics[width=0.95\textwidth]{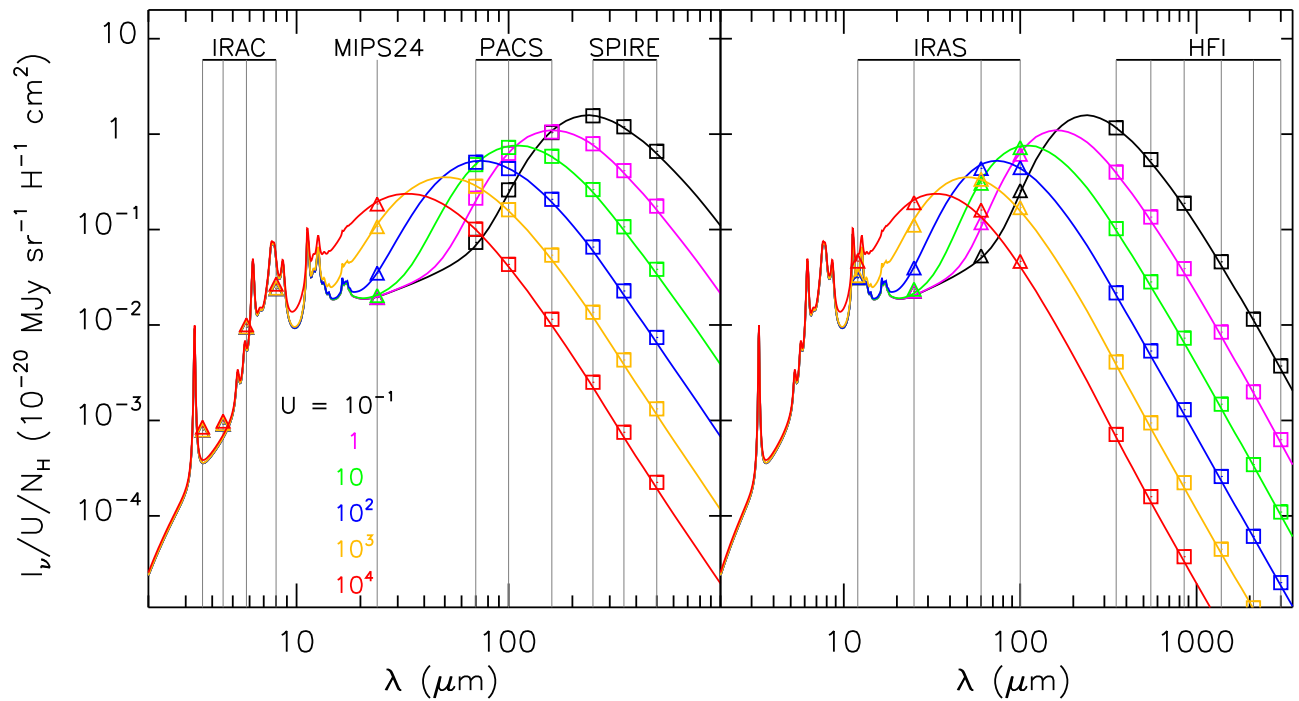}
     \caption{The modeled dust emission per H atom for scaling factors for the 
                    the MMP83 ISRF of $U$=0.1 to 10$^4$.
                    All spectra have been divided by 
                    $U$ to emphasize the changes in the shapes of the SEDs,  which are 
                    presented as $I_{\nu}$ to better indicate the photometric data points.
                    The dust properties are those for DHGL model (see \S\,\ref{sect:reference_model}).
                    The left panel shows the Spitzer (triangles) and Herschel (squares) photometric points.
                    The right panel shows the IRAS (triangles) and Planck/HFI (squares) photometric points. 
                    To emphasize excitation effects we have assumed a constant ionization fraction for the PAHs.
                     }
          \label{fig:emissvschi}
\end{figure*}

To model the DHGL data set we use three dust components: (i) PAHs, (ii) hydrogenated amorphous carbon (hereafter amC) and (iii) amorphous silicates (hereafter aSil).  The properties of each type are described in the Appendix\,\ref{sect:dust_prop}. 
In the course of dust evolution, composite or core-mantle (silicate-carbon) grains are likely to appear that would lead 
to a similar polarization of both the 3.4\,$\mu$m absorption band of amorphous carbon and the 9.7\,$\mu$m band of the silicates \citep[see][]{adamson99,li2002a}.
We do not consider such composite grains here because recent spectropolarimetric measurements by \citet{chiar2006} indicate that the 3.4\,$\mu$m band is not polarized while the 9.7\,$\mu$m band along the same sightline is polarized.
Although such observations should be pursued, these results may indicate that the carriers of the 3.4 and 9.7\,$\mu$m features belong to physically separated dust populations where the former are poorly elongated or not well aligned with the magnetic field while the latter are. Further constraints on this issue will be obtained from Planck measurements of polarized dust emission in the submm range. 
 
The size distribution and abundance of the dust components have been adjusted in order to reproduce both the extinction and emission.  
The required size distributions are shown in the form of their mass distributions in Fig.\ref{fig:sdist}.  
The population of amorphous carbon dust has been divided into small (SamC) and large (LamC) grains, but the overall size distribution of carbonaceous grains (PAH and amC) is continuous. 
Following previous studies \citep[e.g.][]{Weingartner2001}, the smallest grains (PAH and SamC) size distributions are assumed to have log-normal distributions (with $a_0$ the centre radius and $\sigma$ the width of the distribution). The LamC and aSil size distributions follow a power law ($\propto a^{\alpha}$) starting at $a_{\rm min}$ and with an exponential cut-off of the form 
$e^{-[(a-a_t)/a_c]^\gamma }$
for $a\geq a_t$ (1 otherwise) at large sizes. As noted by \citet{kim95}, such a form in the cut-off is required to explain the observed polarization in the near-IR: the polarization capabilities of the present dust model will be discussed in a future paper. The abundances and parameters for the size distributions are summarized in Table\,\ref{tab:model_param}.  PAHs and SamC are small grains of radius less than $\sim10$ nm while LamC and aSil cover a broad size range from a few nm to a few 100 nm. The mean size of PAHs ($a_0\,=\,6.4\,{\rm \AA}$) corresponds to 120 C atoms.
Note that, contrary to carbonaceous dust, for which the PAH bands give a constraint on the abundance and size
of very small particles, only an upper limit can be set regarding the abundance of very small
silicate dust \citep[see][]{li2001a}. In our model, we do not include very small silicate particles.

Our model results are compared to the observational data in Figs.\,\ref{fig:modelemiss}, \ref{fig:modelemiss_lw}, \ref{fig:modelextUV} and \ref{fig:modelext}.  
The dust temperatures were computed using the InterStellar Radiation Field (ISRF) from \citet{mathis83}, hereafter MMP83, for $D_G\,=\,10\,$kpc, labelled ``solar neighborhood''.
Between 5 and 16\,$\mu$m, we convolved the model SED with a Gaussian profile ($\lambda/\Delta \lambda \,\sim\,40$) to account for the ISOCAM/CVF spectral resolution.  The LamC and aSil grains, mostly in thermal equilibrium with the ISRF, emit in the far-IR while PAHs and SamC grains undergo temperature fluctuations and thus contribute to the near- to mid-IR emission. The LamC grains dominate the observed flux at 100 and 140\,$\mu$m since they are hotter than aSil thanks to their higher absorption efficiency at $\lambda \sim 1\, \mu$m (see Fig.\,\ref{fig:nk}) where the MMP83 ISRF peaks.  We note that the SamC grains make a significant contribution at 25 and 60\,$\mu$m. At $\lambda \ga 250\,\mu$m, the aSil and LamC contribute almost equally to the SED down to the WMAP measurements that are well reproduced (see Fig.\ref{fig:modelemiss_lw}) denoting no significant variation in the spectral emissivity index between $\sim250\,\mu$m and $\sim$\,5\,mm. At these wavelengths $\beta\sim1.5-1.6$ for both of the amC and aSil materials.

The IR extinction (Fig.\,\ref{fig:modelext}) is dominated by the contribution of the LamC and aSil grains.  The 9.7 and 18\,$\mu$m bands are produced by silicates while the 1\,-\,10$\,\mu$m continuum is due to amorphous carbon as seen in Figs.\,\ref{fig:modelext} and \ref{fig:nk}. In the DL07 model, these two spectral components are also produced by amorphous silicate and big carbonaceous (graphite) grains.  The 217 nm extinction bump (see Fig.\,\ref{fig:modelextUV}) is entirely due to aromatic $\pi\,\rightarrow\,\pi^\star$ transitions in PAHs, as proposed by \citet{iati2008} and \citet{Cecchi-Pestellini2008} and supported by the experimental results of \citet{steglich2010}.  The far-UV extinction rise is also mostly produced by PAHs. This latter feature is common to most carbonaceous grains: it is the red wing of an absorption peak around 750\,\AA\ due to $\sigma\,\rightarrow\,\sigma^\star$ transitions \citep[][]{jager2009}.  In our model, the bump strength and the PAH emission are thus correlated. We note that \citet{boulanger94} found a trend between the bump area and the IRAS 12 and 25\,$\mu$m emission, suggesting that small amorphous carbon grains may also contribute to the bump \citep[see also][]{mathis94}.  The 217\,nm and 750\,\AA\ broad bands are quite weak for the BE amC sample that we used.  We emphasize that using a carbonaceous material with a strong 217\,nm band (e.g. graphite) would have resulted in an overestimation of this feature in our model, given the abundances required for both PAH and the small carbonaceous grains to reproduce the Aromatic Infrared Bands (AIB) and mid-IR continuum.  Indeed, in the DL07 model, PAHs account for 4.6\% of the total dust mass compared to 7.7\% for our model) which allows for about half of the 217\,nm bump to be produced by small graphite particles.  This directly translates into a difference of a factor of $\sim2$ in the AIB emission intensity between the two models as shown on Fig.\ref{fig:compare_emiss}.

The dust in our model, excited by the MMP83 ISRF, radiates a total power of $4\pi \int \nu I_\nu d\nu\,/N_H=\,5.0\times 10^{-24}$ erg s$^{-1}$ H$^{-1}$ with a fractional contribution of 29, 7, 38 and 26\% for PAH, SamC, LamC and aSil, respectively. This value for the total emitted power by dust is intermediate between those of DBP90 and DL07 (U=1, $q_{PAH}\,=\,4.6\,\%$), $6.9\times10^{-24}$ erg s$^{-1}$ H$^{-1}$ and $4.5\times10^{-24}$ erg s$^{-1}$ H$^{-1}$, respectively, while the three models have about the same extinction curve (the observed one). We note that in the case of DBP90, the reference SED (to which the model is fit) was higher as can be seen in Fig.\ref{fig:compare_emiss} implying a higher dust absorption efficiency. Indeed, we see in the bottom panel of Fig.\,\ref{fig:modelextUV} that the DPB90 model has the lowest albedo followed by our model, while the DL07 model has the highest albedo, thus explaining the differences in emitted power noted above. As already discussed above, the difference between our model and the DL07 modeled SEDs mostly resides in the AIB spectrum intensity while the two models agree rather well at $\lambda\ga60\,\mu$m.

The total dust-to-gas mass ratio in our model is $M_{dust}/M_H\sim10.2\times10^{-3}$, corresponding to $M_{gas}/M_{dust}\sim133$ (with $M_{gas}/M_H\,=\,1.36$). This value is equal to the value for the DL07 model ($M_{dust}/M_H\sim10.4\times10^{-3}$) and higher than the DBP90 model ($M_{dust}/M_H $ of $7.3 \times 10^{-3}$).  
The required elemental abundances are in agreement with measurements (see  Table\,\ref{tab:abundances}), except for iron for which our model requires $\sim30\%$ more than 
the values inferred from both F,G stars and the Sun.
The carbon is shared between PAH, SamC and LamC with 32, 7 and 61\% in each respectively. 
Our model requires about the same amount of carbonaceous material and silicate as the DL07 model (234\,ppm for C, compared to 200\,ppm for our model, and 44\,ppm of MgFeSiO$_4$ for DL07 compared to 45\,ppm for our model).
In our model, silicate dust represents $\sim76\%$ ($\sim$73\% in DL07) of the total dust mass.
This mass ratio depends on the assumed densities then on the assumed form for both carbonaceous and silicate material.
On the other hand, the dust material is assumed to be compact to keep the model as simple as possible but considering 
porosity would induce a change in the carbon/silicate mass ratio
(mainly due to change of optical properties)
and decreases the required abundances.

\begin{figure}
    \centering
      \includegraphics[width=0.49\textwidth]{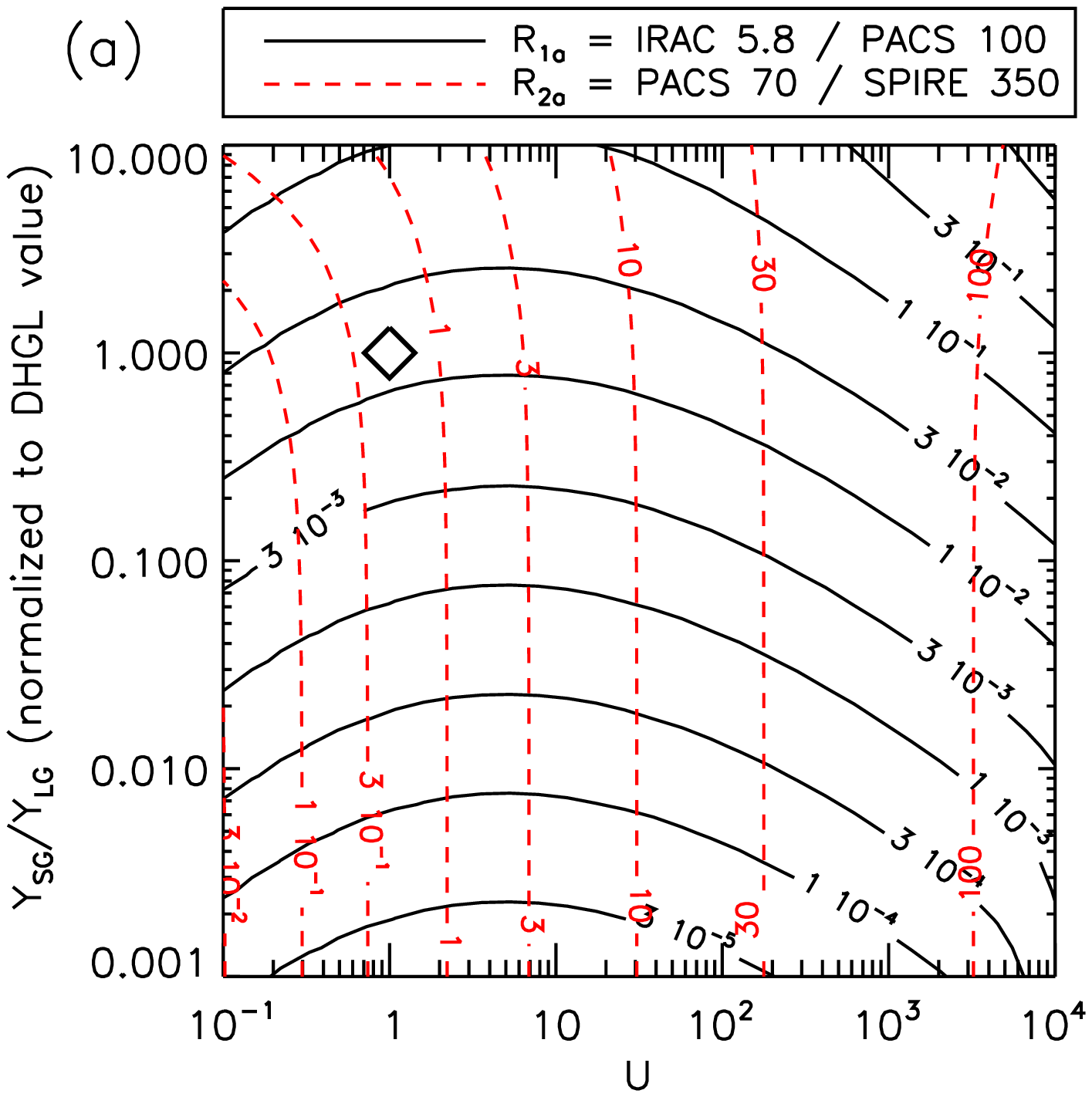}
      \includegraphics[width=0.49\textwidth]{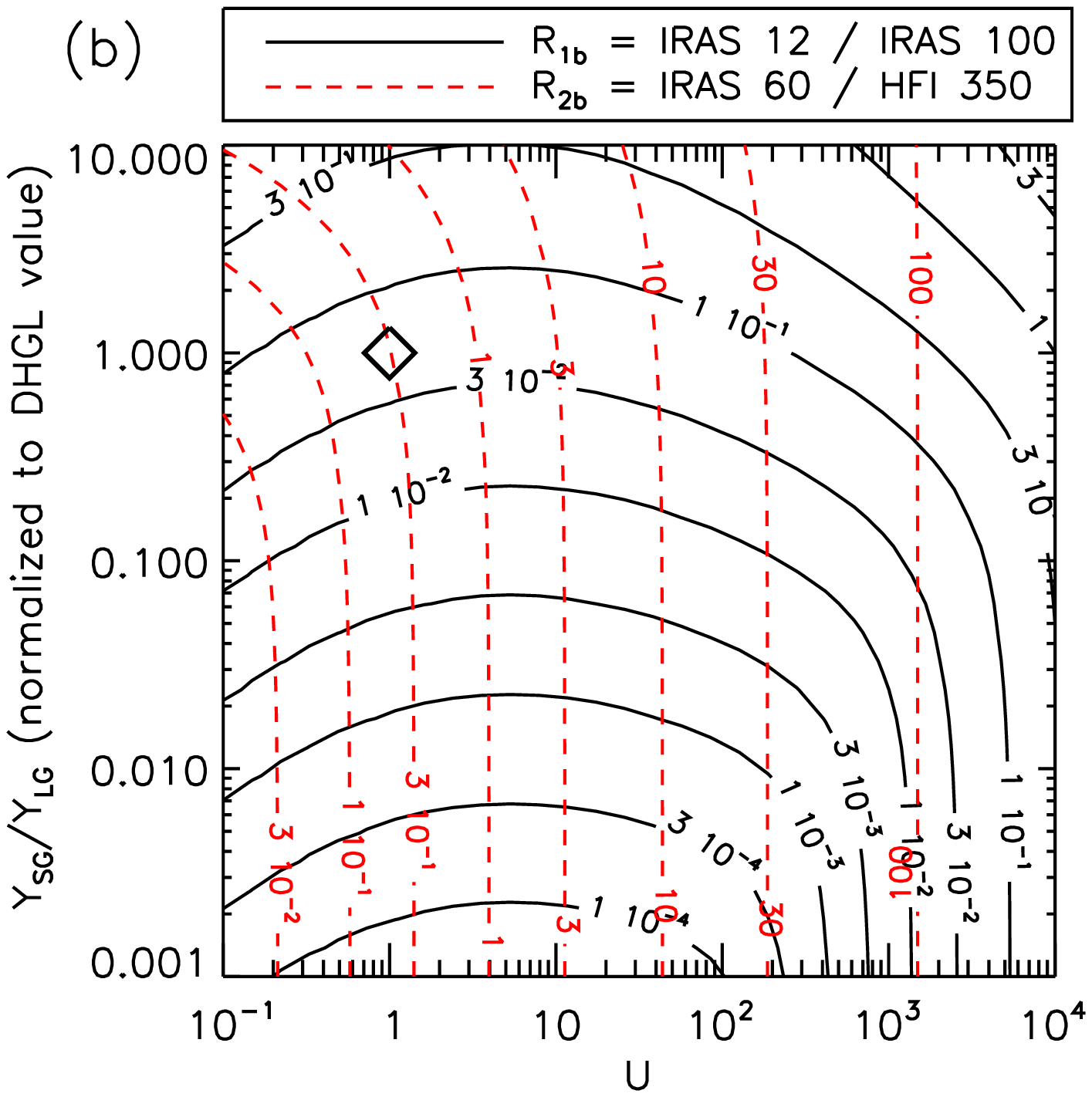}
    \caption{
              The predicted photometric band ratios $R_1$ (mid-IR) and $R_2$ (far-IR) as a function of $U$ 
              and the ratio between the mass abundance of small
              ($Y_{SG}$) and large ($Y_{LG}$) grains (see text). 
               Note that the band ratios are shown on logarithmic scales.
              $Y_{SG}\,/\,Y_{LG}$ is given relative to the diffuse interstellar medium value of 0.10 for
              the DHGL value, which is shown by the open diamond. 
             }
          \label{fig:figratios}
\end{figure}
 
\section{Tracing the nature and evolution of dust using a global SED}
\label{sect:evolution}

\begin{figure*}[t]
    \centering
      \includegraphics[width=1.00\textwidth]{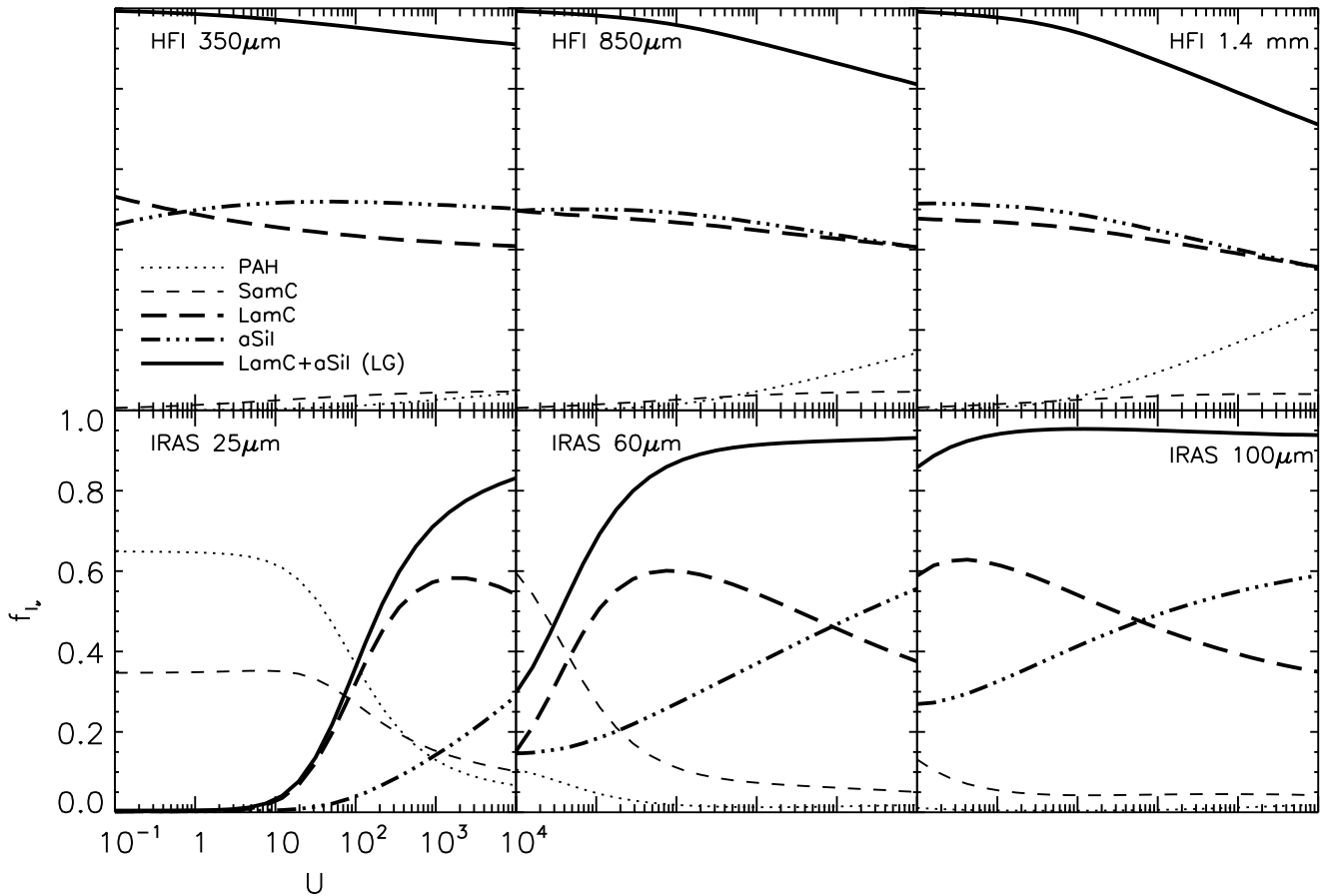}
     \caption{The fractional contributions of the dust populations in our model to the IRAS and Planck photometric 
              bands as a function of $U$.
              The bold lines emphasize the contribution of
              the large grains, LamC and aSil. 
              The dust properties are those for DHGL model (see \S\,\ref{sect:reference_model}).
}
          \label{fig:fractionInu}
\end{figure*}

\begin{figure*}[t]
    \centering
      \includegraphics[width=1.00\textwidth]{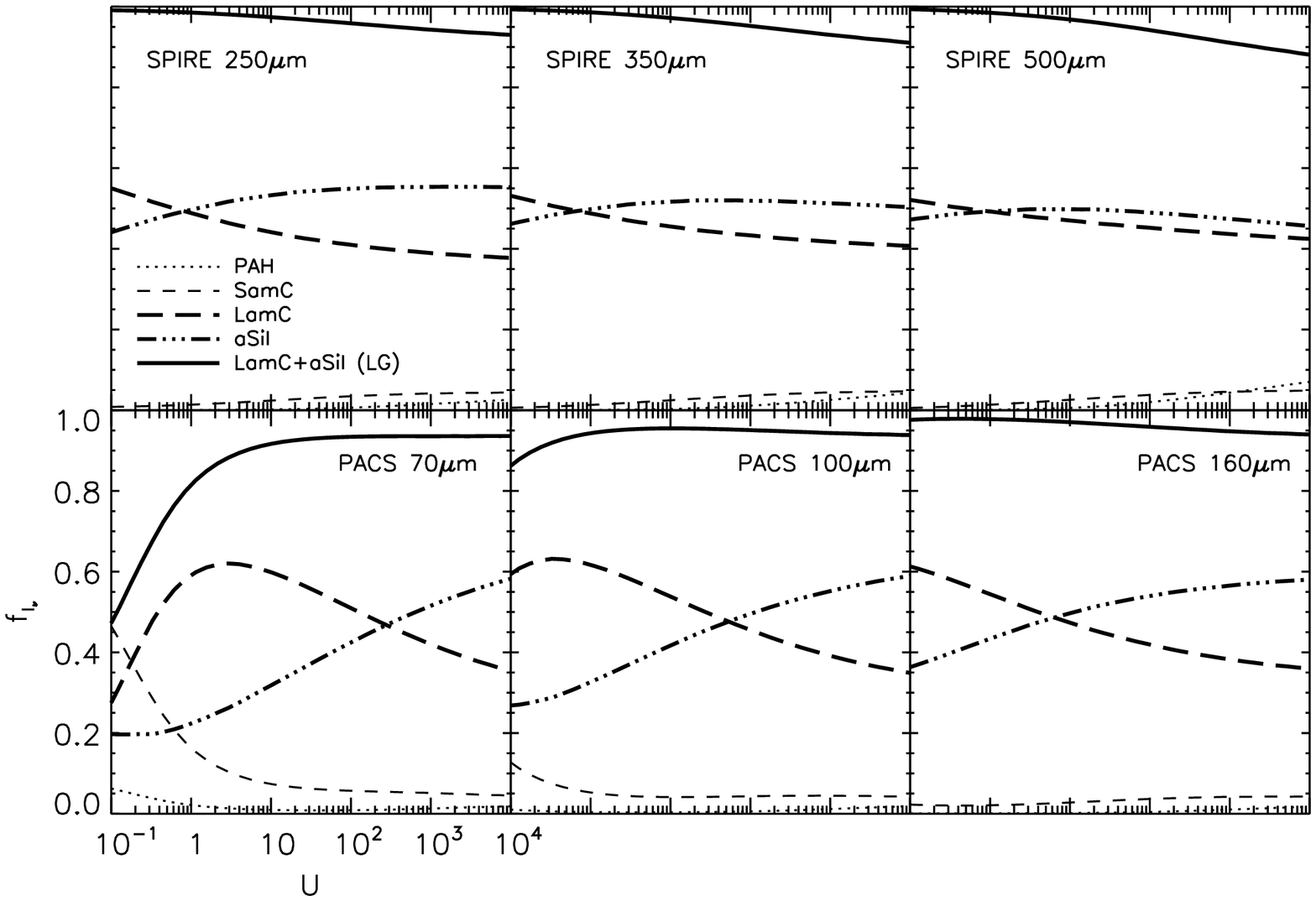}
     \caption{Same as Fig.\,\ref{fig:fractionInu} but for {\it Herschel} SPIRE and PACS bands.}
          \label{fig:fractionInu_pacsspire}
\end{figure*}

Thanks to the Herschel and Planck missions we will soon have at our disposal the global dust SED, by combining these data with the already available near- and mid-IR data at comparable angular resolution (IRAS, ISO and Spitzer).  The availability of this global dust SED will allow us, for the first time, to make a systematic study of dust evolution throughout the ISM. Indeed, dust evolution processes (e.g. coagulation, fragmentation) primarily affect the dust size distribution and the abundance ratios of the various dust components and are therefore directly reflected in the observed dust SED, which is also affected by the intensity of the ISRF.  We first examine the effects of the latter on the model dust emission and the behavior of the associated fluxes in the photometric bands.  For this purpose we scale the MMP83 ISRF by a factor, $U$, using the same nomenclature as DL07.  Note that as we scale the entire exciting ISRF, we have $U\,=\,G_0$, where $G_0$ is the ratio of the intensity of the radiation field integrated between 6 and 13.6 eV relative to the standard ISRF.

Fig.\,\ref{fig:emissvschi} shows the dust SED for ISRF intensities ranging from $U\,=\,10^{-1}$ to 10$^4$.  We overlay the corresponding photometric points for several instruments and have separated the high (left panel) and low (right panel) spatial resolution instruments. 
We do not show the MIPS 70 and 160\,$\mu$m bands that fall at the same wavelengths as the Herschel/PACS bands.  To illustrate changes in the SED shape, the emission flux has been divided by $U$. As expected, we see that the emission of PAHs, which undergo temperature fluctuations, scales linearly with $U$. By contrast, the emission from the large grains (at thermal equilibrium with the ISRF) shows a more complex behavior, scaling but also shifting to higher frequencies as $U$ increases.  As $U$ increases smaller and smaller grains reach thermal equilibrium with the ISRF (see for instance, Fig.4 of DL07) and consequently follow the same type of behavior as the largest grains.  This is the reason why the SamC population shows intermediate behavior in the presented $U$ range.  
The general behavior of our model regarding $U$ is similar to that of DL07.
However, a significant difference arises at $U\geq 10^{3}$ due to the fact that we use amorphous carbon rather than graphite.
At these values of $U$ our large grain (LamC and aSil) spectrum is remarkably broader, because we lack the broad feature in the absorption (i.e. emissivity) efficiency of graphite at $\lambda\sim30\,\mu$m introduced by DL07.

Here we illustrate how the effects of the intensity of the ISRF can be disentangled from effects of the relative abundances of small-to-large grains (i.e., PAH+SamC to LamC+aSil).  The latter variation is equivalent to a changing size distribution, and is to be expected when small grains coagulate onto, or disaggregate from, larger grains. We take $Y_i$ to be the mass abundances of dust population $i$ (see Tab.\ref{tab:model_param}) and define $Y_{SG}\,=\,Y_{PAH}\,+\,Y_{SamC}$ and $Y_{LG}\,=\,Y_{LamC}\,+\,Y_{aSil}$, the mass abundances of small grains (SG) and large grains (LG), respectively. Note, that while varying $Y_{SG}$, we keep $Y_{LG}$ constant and, for simplicity, assume that the opacity of the large grains is unchanged even though it should change when coagulation occurs\footnote{This process has been shown to alter the sub-mm opacity by factors 2 to 3 \citep{stepnik2003,paradis2009b}.}.  We also note that, even though the PAH/SamC abundance ratio is taken to be constant for this illustrative case, the relative abundance between the PAH band emitters and the mid-IR continuum emitters (SamC in our case) is known to vary \citep[e.g][]{berne2007, compiegne2008, flagey2009}.
Fig.\,\ref{fig:figratios} shows the photometric ratios $R_{1a}=IRAC\,5.8/PACS\,100$, $R_{2a}=PACS\,70/SPIRE\,350$ (top panel) and $R_{1b}=IRAS\,12/IRAS\,100$, $R_{2b}=IRAS\,60/HFI\,350$ (lower panel) as a function of $U$ and $Y_{SG}/Y_{LG}$. The photometric fluxes were obtained by integrating the spectra over the relevant filter transmission taking into account the color corrections. The ratios $R_1$ and $R_2$ represent roughly  the same wavelengths in case (a) and (b) and so consequently show a similar behavior.  We see that for $10^{-3}<Y_{SG}/Y_{LG}<10$ and $10^{-1}<U<10^{4}$, the band ratios $R_1$ (a and b) scale with $Y_{SG}/Y_{LG}$ because the SG emission becomes dominant at small wavelengths. Conversely, the $R_2$ ratios (a and b) are mostly sensitive to variations  in $U$ because the photometric bands involved are dominated by the LG contribution (see Fig.\,\ref{fig:fractionInu} and \ref{fig:fractionInu_pacsspire}).  In this range of $U$ and $Y_{SG}/Y_{LG}$ parameter space the variations induce well discriminated differences in the photometric band ratios $R_i$.  From the observed values of $R_1$ and $R_2$, $U$ and $Y_{SG}/Y_{LG}$ can thus be constrained for a given dust model. However, we note that for high $U$ and low $Y_{SG}/Y_{LG}$ $R_{1b}$ and $R_{2b}$  this discrimination is not possible because the spectrum is dominated by BGs down to wavelengths as short as 12\,$\mu m$. This degeneracy is avoided for higher $U$ when using the shorter wavelength to trace SG emission, as seen with $R_{1a}$.

In general, a photometric point measured on the dust SED is a sum of the contributions of the different dust components (composition and sizes) and the analysis of photometric data therefore requires prior knowledge of these contributions. In view of the differing behaviors of the SG and LG dust emissions, with respect to $U$ (Fig.\,\ref{fig:emissvschi}), we show in Fig.\,\ref{fig:fractionInu} and \ref{fig:fractionInu_pacsspire} the fractional contributions of the model dust populations (for the DHGL properties) in several photometric bands, as a function of $U$. For $10^{-1}<U<10^{4}$, in the IRAS\,$25\mu m$ band (Fig.\,\ref{fig:fractionInu}), the LG contribution gradually increases as $U$ increases because these grains become warmer, as can be seen in Fig.\,\ref{fig:emissvschi}. The LamC contribution is dominant (with respect to aSil) at $\lambda\la60\,\mu m$ because these grains are warmer, a consequence of their higher absorptivity in the near-IR (Fig.\,\ref{fig:nk}). At $\lambda\sim70-100\,\mu m$ ($\lambda\sim160-250\,\mu m$), and for this same reason, LamC dominates at $U\la10^{2}$ ($U\la5$) while aSil does only for higher $U$ since the LamC grains then emit at shorter wavelengths.  LamC and aSil have about the same contribution at $\lambda\ga 350\,\mu$m (as seen for $U=1$ in Fig.\ref{fig:modelemiss}).  While the wavelength of the LG emission peak moves to higher frequencies when $U$ increases ($U\ga 10^2-10^3$), the cold part of the temperature fluctuations in PAHs contributes to the emission in the HFI 1.4 mm (217 GHz) band (Fig.\,\ref{fig:fractionInu}). We note that at these frequencies recent theoretical work indicate that rotational emission of PAHs for $G_0\ga 10^3$ and gas densities $n_H \sim 10^5$ cm$^{-3}$ might also contribute \citep{hoang2010,silsbee2010}.  We emphasize the fact that the dust contributions at 60 and 100\,$\mu$m allow for a better determination of the LG SED around its emission peak and hence better mean temperature estimates for the LGs, a necessary step in the extraction of the thermal dust component in Planck data.  Finally, and as discussed earlier, the emission of aSil grains is expected to be polarized unlike that of the LamC grains and PAHs \citep{sironi2009}. A determination of the respective contributions of the different LGs and of PAHs to the dust SED will therefore be crucial to the analysis of data from the Planck-HFI polarized channels (at $\lambda>850\,\mu m$).

\section{Summary}
\label{sect:summary}
The emission from large interstellar grains ($a\sim 0.1\;\mu$m) is currently being observed by the instruments onboard the Planck and Herschel satellites. The properties of the bulk of this dust are likely determined by dust evolution throughout the lifecycle of interstellar matter and are probably not just a reflection of the composition of unaltered `stardust'. The primary effect of dust evolution is to redistribute dust mass during growth or fragmentation episodes, resulting in changes in the abundance ratio of small-to-large grains, $Y_{SG}/Y_{LG}$, (where small grains have $a \la 10$ nm). Such changes are reflected in the global (IR to mm) dust SED, which will soon available for large parts of the sky, thanks to the combination of Spitzer, IRAS, Herschel and Planck data. Understanding dust evolution is critical for the physics and chemistry of the ISM but also for the optimum subtraction of the thermal dust contribution from the  Planck data in CMB studies.

Dust evolves in the ISM  and thus shows regional variations in its intrinsic properties, structures and sizes. In this work we describe a framework dust model and provide the tools necessary for the study of dust evolution in the ISM.  We present a versatile numerical tool, {\tt DustEM} , that allows us to predict the dust SED and extinction, given the dust properties, and that is capable of handling a large diversity of dust types, using a knowledge of their size-dependent opacities and heat capacities. The design of {\tt DustEM}  allows the user to add new grain physics in a simple way. To enable an assessment of the impact of dust evolution we have defined a reference dust model for the DHGL medium, where the dust populations have certainly undergone growth and fragmentation processing and where the calculated dust SED and extinction represent the `average' or `equilibrium' observables for evolved dust.  Our dust model comprises (i)\,Polycyclic Aromatic Hydrocarbons, (ii)\,amorphous carbons and (iii)\,amorphous silicates. We argue that hydrogenated amorphous carbon grains are the natural product of dust evolution in the ISM, as are amorphous silicates. Although composite grains may naturally arise in the ISM, we did not include them here, on the basis of the results from recent spectropolarimetric observations. The Planck polarized channel data will soon make available more constraints in this issue. 

The dust SED reflects changes in the abundances and sizes of grain populations and also variations in the intensity of the exciting radiation field. Using {\tt DustEM}  we have shown that well chosen photometric band ratios can disentangle the influence of the intensity of the exciting radiation field from that of changes in dust abundances as indicated by $Y_{SG}/Y_{LG}$. We show the contributions of the small and large grains in the IRAS and Planck-HFI channels (and similarly to Herschel channels). Such information allows for a reliable determination of the temperature of large grains, a necessary step in the quantitative subtraction of thermal dust emission from the Planck data. These results assume fixed dust optical properties and size distributions. However, in a realistic treatment of dust evolution such quantities will change with the local physical conditions and affect the FIR to sub-mm grain emissivity. Future work will require the use of IR, Herschel and Planck data to assess variations of $Y_{SG}/Y_{LG}$ in interstellar regions where small grains may coagulate onto, or result from the fragmentation of, larger grains, resulting in changes in the sub-mm opacity of dust. Combined with the constraints from Planck polarized channel data, these full SED dust data should provide plausible constraints on the candidate materials, and their physical properties, that can be invoked for the evolved dust in the ISM.

\begin{acknowledgements}
We thank A. Abergel, V. Guillet and M.-A. Miville-Desch\^enes for stimulating discussions.  We are also grateful to V. Guillet for providing us with his routine yielding the ionization fraction of PAHs. This research acknowledges the support of the french ANR through the program {\it Cold dust} (ANR-07-BLAN-0364-01).
\end{acknowledgements}

\begin{appendix}
\section{Dust properties}\label{sect:dust_prop}

\begin{figure*}[t]
   \centering
     \includegraphics[width=0.49\textwidth]{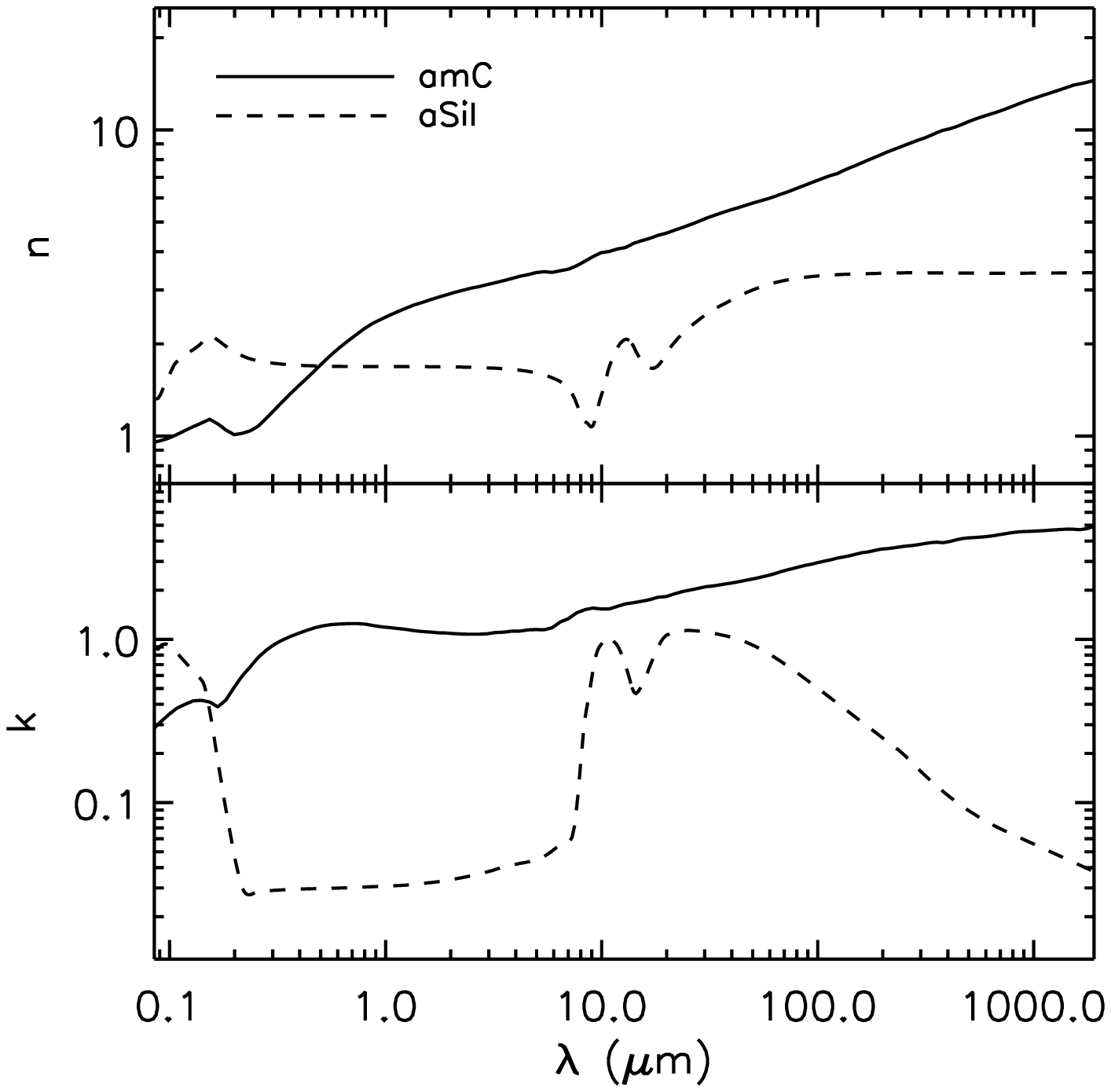}
     \includegraphics[width=0.49\textwidth]{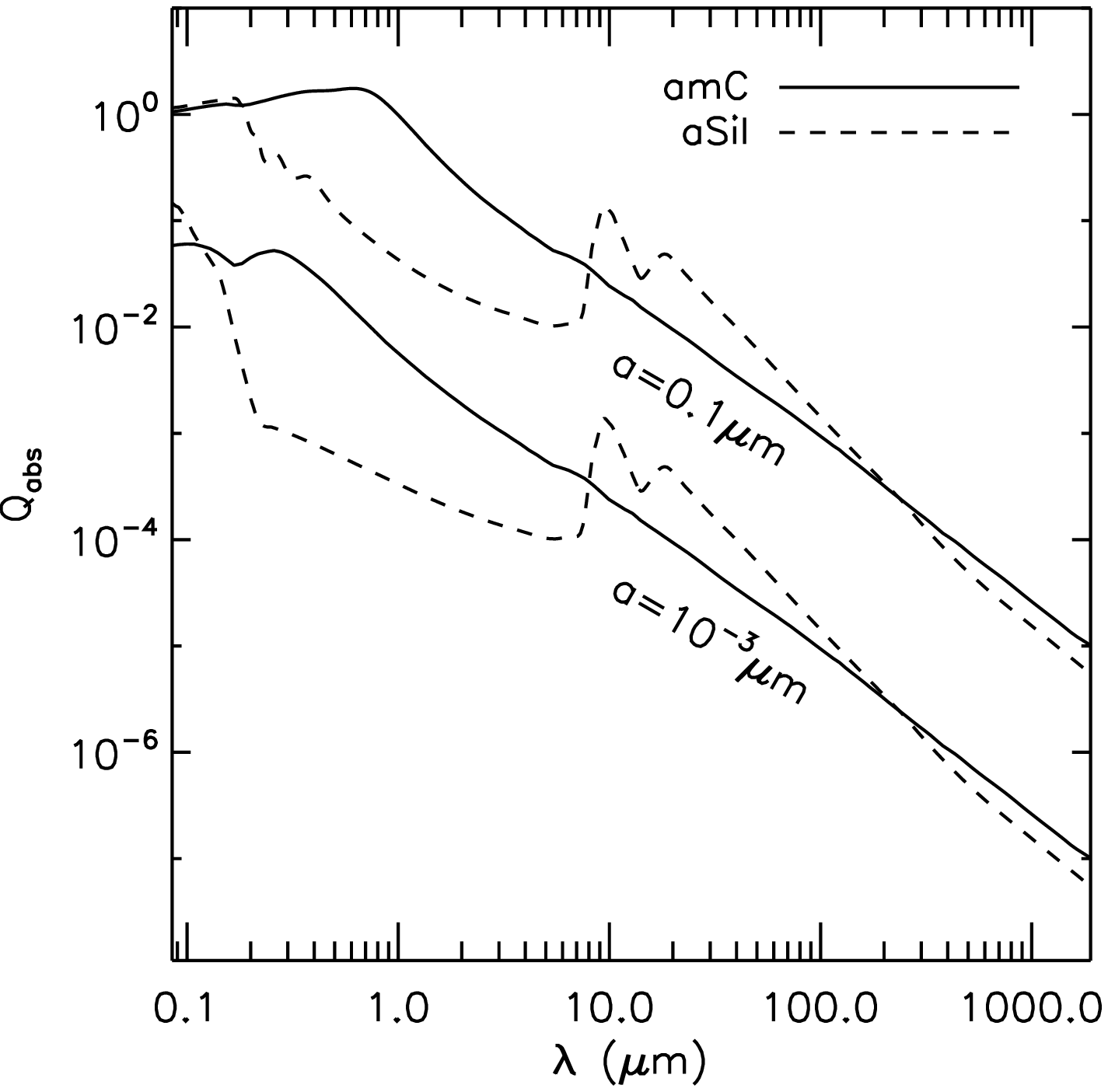}
      \caption{The refractive indices ($m\,=\,n+ik$) of amorphous carbon and
                    amorphous silicates (left panel) and the corresponding
                    absorption efficiencies obtained from a Mie
                    calculation (right panel).
                   }
         \label{fig:nk}
\end{figure*}

Here we discuss the optical and thermal properties of the grain types adopted in this work. 

\subsection{PAHs}
\label{sect:PAH}

Although a global picture of the contributions of different PAH families to the mid-IR bands is emerging (e.g. \citet{bauschlicher2009}), it does not yet quantitatively describe how the band profiles vary as a function of the physical conditions (radiation field intensity and gas density). As in the case of silicates (see hereafter in the Appendix), an empirical approach can be followed where the ratios and profiles of the IR bands are derived from observations \citep{verstraete2001,li2001}. This approach is however degenerate, given the fact that the band ratios depend on size, ionization and hydrogenation states. Recently, \citet{draine2007} (hereafter DL07) added the constraints provided by the extensive theoretical database of \citet{malloci2007} to define a new set of band parameters, both for neutral and cationic PAHs. Here we use the cross-sections defined in DL07 with the following modifications. To allow for a better match to the ISOCAM/CVF spectrum of the diffuse interstellar medium, presented in \citet{flagey2006} and in Fig.\,\ref{fig:modelemiss}, we increase the integrated cross-sections $\sigma_{int}$ of the 8.6, 11.3 and 12.7 $\mu$m bands by 50\%, 10\% and 50\%, respectively.  We have not included the near-IR bands between 1 and 2 $\mu$m, which have little effect on the emission. We added the far-IR bands defined by \citet{ysard2010} from the Malloci database. The PAHs are assumed to be disk-like and their size, $a$, is defined to be that of an equivalent sphere of the same mass, assuming a density of $\rho=2.24$ g/cm$^3$. The relation between $a$ and $N_C$ is $N_C\simeq 470\,\left(a/1{\rm nm}\right)$.  We assume the  PAHs to be fully hydrogenated, with [H/C] varying as a function of $N_C$ as defined in DL07. Following DL07, the same UV-visible absorption cross-section is assumed for neutral and charged PAHs while the same absorption cross-section is assumed for cations and anions.  The size dependence of the ionization fraction of PAHs is also obtained in the same way as by DL07: it has been estimated using the \citet{weingartner2001c} formalism and for the physical conditions summarized in \citet{li2001} (see their Table\,2) for the DHGL medium.  The heat capacity has been taken from \citet{Draine2001}. 

\subsection{Amorphous carbon}
\label{sect:amC}

Amorphous carbons can assume many different structures and compositions, depending on the way in which the sp$^2$ and sp$^3$ carbon atoms combine with one another and with other species, especially with hydrogen.  \citet{zubko96a} compiled the optical properties of several solid hydrocarbons over a broad spectral range, from the millimeter to the EUV range. These optical properties were derived from laboratory measurements of the opacity made by \citet{colangeli95}. In this work we adopt the optical properties of the BE sample: formed by benzene combustion in air, this sample is representative of rather sp$^2$-rich hydrogenated amorphous carbon.  The refractive index for the BE material is displayed in Fig.\,\ref{fig:nk} (left).  The absorption and extinction efficiencies were obtained from Mie calculations. Fig.\,\ref{fig:nk} (right) shows the absorption efficiencies, $Q_{abs}$, for grains of  radius 1 and 100 nm. The refractive index given by \citet{zubko96a} goes down to $\sim$1900$\mu m$. We extrapolated the optical efficiencies down to 10\,cm using the average emissivity index between 800 and 1900\,$\mu m$, using $\beta\,=\,1.55$. This $\beta$ value does not depend on the size (for $a\la3\,\mu m$).
For the heat capacity \citet{michelsen2008} argue that the graphite heat capacity is a good surrogate for soots and for amorphous carbons: we therefore use the graphite heat capacity as defined in \citet{Draine2001}. Finally, we adopt a mass density $\rho\,=\,1.81$ g cm$^{-3}$, typical of hydrogenated amorphous carbons.

\subsection{Astronomical silicate}
\label{sect:aSil}

The observed profiles of the 9.7 and 18 $\mu$m absorption bands indicate that the interstellar amorphous silicates are of olivine-type (Mg$_x$Fe$_y$Si0$_4$). The laboratory band profiles are, however, too narrow to explain the observations, suggesting that some kind of processing is at work and that other types of silicates contribute to these bands \citep{demyk2004}. \citet{draine84} followed an empirical approach and defined astronomical silicates whose optical properties in the mid-IR were constructed from observational data. \citet{li2001} further modified the imaginary part of the dielectric function for $\lambda \geq 250\;\mu$m to better match the high galactic latitude dust emission measured by FIRAS. Using the optical properties described in \cite{draine2003a}\footnote{{\tiny Available at \\ {\tt http://www.astro.princeton.edu/$\sim$draine/dust/dust.html}}}, 
we derived the absorption and extinction efficiencies using Mie calculations. The right part of Fig.\,\ref{fig:nk} shows the corresponding absorption efficiencies, $Q_{abs}$.  Finally, we take the heat capacity for amorphous silicates from \citet{Draine2001} and assume a mass density of 3.5 g cm$^{-3}$.

\section{The temperature-dependent grain emissivities tool}
\label{sect:qabstemp}
In this section we present a tool that allows the user to apply a change in the long wavelength opacity spectral index ($\beta$) as a function of the wavelength and the temperature.

The determination of dust temperatures and the dust spectral emissivity index, $\beta$  ($Q_{abs}\sim\nu^{\beta}$), from far-IR to submm data is degenerate and sensitive to the errors or multiple dust temperatures \citep{dupac2003,shetty2009a,shetty2009b}. Nevertheless, observations do seem to indicate a temperature-dependence of $\beta$ \citep{dupac2003,desert2008}, which may reflect the microscopic properties of grains at low temperatures \citep{meny2007}. In addition, laboratory data on the sub-mm opacity of interstellar dust analogues show a similar temperature dependence \citep{agladze1996, mennella1998,boudet2005}. 
Within the perspective of the Herschel/Planck data interpretation, the possibility of temperature-dependent dust emissivities has been introduced into {\tt DustEM} . This feature is implemented as a correction to the 
emissivity index $\beta$ and takes the form:
\begin{equation}
Q_{abs}(a,\nu) = Q_0(a,\nu)\,\left({\nu/\nu_t}\right)^{\delta(T)\,H(\nu_t/\nu)}
\end{equation}
where $Q_0$ is the grain emissivity without the temperature dependent index, $\nu_t$ is the frequency threshold below which the correction is applied, $\delta(T)$ is the index correction and $H$ is the threshold function. We define the correction as $\delta(T)=\beta(T)-\beta_0$ where $\beta(T)$ is the temperature law for the emissivity index and $\beta_0$ is some reference value.  Such a $\beta$-correction should then preferentially be applied to grain types whose $Q_{abs}$ behave as a power laws (with an index $\beta_0$) at frequencies lower than the threshold,  otherwise $\beta_0$ may depend on the frequency.
The $\beta(T)$-values are estimated with a user-defined, parameterized function or can be interpolated in a data table. The parameters, as well as the tabulated values (if used), are read from a data file.  We chose the threshold function to be $H(x)=(1+4s\tanh(x-1))/2$ where $s$ controls the stiffness of the transition from 0 to 1 ($\Delta x = x/s$): for instance for $s=1$ the $H$-values rises from 0 to 1 for $x$ between 0.5 to 1.5 and the width of the transition is $\Delta x=1$.  Since the temperature dependence of the dust emissivity appears at long wavelengths ($\lambda\geq 30\;\mu$m), where the dust absorption efficiency is much smaller than in the visible and near-IR range (see Fig.\,\ref{fig:modelext}), we apply the $\beta$-correction only in the cooling function (emission) for the dust in the {\tt DustEM}  code.
\end{appendix}

\bibliographystyle{aa} 
\bibliography{15292}

\end{document}